\documentclass[letterpaper,11pt]{article}
\usepackage[british]{babel}
\usepackage[latin1]{inputenc}
\usepackage{graphicx}
\usepackage{epsfig}
\usepackage{amssymb}
\usepackage{amsmath}
\usepackage{amsfonts}

\usepackage[font=small,labelfont=bf]{caption}

\newcommand{\id}{{\hbox{{\rm 1}\kern-.24em\hbox{\rm l}}}}

\begin{document}

\vspace*{10mm}
\begin{center}
  {\LARGE \bf{Properties of QBist State Spaces}} \\[8mm]

D. M. Appleby,
\AA sa Ericsson,
and Christopher A. Fuchs
 \\[2mm]

{\small\sl Perimeter Institute for Theoretical Physics} \\
{\small\sl 31 Caroline Street North, Waterloo, Ontario N2L 2Y5, Canada} \\[10mm]

{\bf Abstract} \\[4mm]

\begin{minipage}{0.9\linewidth}
\hspace{4mm}{\small
Every quantum state can be represented as a probability distribution over the outcomes of an informationally complete measurement. But not all probability distributions correspond to quantum states. Quantum state space may thus be thought of as a restricted subset of all potentially available probabilities. A recent publication~\cite{Fuchs&Schack} advocates such a representation using symmetric informationally complete (SIC) measurements. Building upon this work we study how this subset---quantum-state space---might be characterized. Our leading characteristic is that the inner products of the probabilities are bounded, a simple condition with nontrivial consequences. To get quantum-state space something more detailed about the extreme points is needed. No definitive characterization is reached, but we see several new interesting features over those in~\cite{Fuchs&Schack}, and all in conformity with quantum theory.}
\end{minipage}
\end{center}

\vspace{2mm}

\begin{center}\vspace{2mm}
  {\Large \bf I. \ Introduction}
\end{center}\vspace{2mm}

When striving to grasp the meaning of quantum theory, an essential issue is to understand its space of states.  What is quantum-state space, and how does it compare to classical state space and other possible theories?

An example of an approach widely used to compare quantum states with classical states is to study the Wigner-function representation of the former. Quantum states are almost like probability distributions over a classical phase space, but the catch is that these functions can be negative. With the purpose of differentiating quantum theory from other theories---classical, or any of a wide variety of other strange creatures---another approach has got a lot of attention lately. This is the convex operational framework~\cite{Barrett07,Barnum06,Barnum08}. From general considerations about measurements and measurement outcomes (and the probabilities for those outcomes) one derives that the state space of a theory is a convex set and measurements are related to the cone dual to the cone having the state space as its base. The motivation for these studies comes from quantum information theory with its no-cloning theorem, secret key distribution, teleportation, no bit-commitment and more. Within the convex-sets framework one can consider more general theories and ask to what extent ``quantum'' features appear in these as well.  In the convex sets framework classical theory corresponds to a simplex, the simplex of all probabilities, and the convex set of quantum states is just the ordinary set of density matrices, that is, all positive-semidefinite unit-trace operators on some complex Hilbert space.

We will consider a similar kind of representation of quantum states recently put forward by one of us, Fuchs, together with Schack~(F\&S hereafter)~\cite{Fuchs&Schack}. See also the contribution by Fuchs and Schack in this special issue~\cite{Fuchs&SchackII}. By use of a special measurement---a symmetric informationally complete (SIC) measurement---the set of density matrices in finite dimensions can be mapped uniquely to a set of probability distributions over a fixed number of outcomes. In this way, the set of quantum states can be seen as a convex set of probability distributions, with a {\it single\/} probability distribution corresponding to each state. But not all probability distributions correspond to quantum states---quantum-state space is a subset of the probability simplex.  From this point of view, quantum-state space is a restriction of the standard probabilistic case---namely, it is a statistical model~\cite{Sullivant}---as opposed to the common view that quantum theory is a noncommuting {\it generalization\/} of probability theory. In the following we will study how this restriction may be characterized. We will use inequalities that depend on the Born rule in an interesting way. In this setting, the Born rule concerns how to assign probabilities to the outcomes of one measurement when given the probabilities for the outcomes of another, counterfactual, measurement. The requirement that all the probabilities really are probabilities---that is, that they are positive and sum to one---gives a restriction that tells us at least part of the story of what quantum-state space looks like.

Additional criteria include requiring a special basis of distributions and a statement about extreme points. These are either proposed or hinted at in F\&S~\cite{Fuchs&Schack}. This paper extends that work, demonstrating a few new properties for state spaces of this variety. We also provide some alternative, clarified proofs for some of the previous results in~\cite{Fuchs&Schack}. We call our state spaces ``QBist'', alluding to the term ``QBism'' coined by F\&S for the quantum-Bayesianism interpretation of quantum mechanics they are developing. The point of view taken seriously in their approach is that all probabilities, thus also quantum states, are personal in the Bayesian sense.

In the following section the SIC-representation of quantum states is reviewed. Section~III is the main part of the paper, where we explore aspects of the characterization of QBist state spaces.  Finally in Section~IV we briefly discuss where things stand.

\begin{center}\vspace{2mm}
  {\Large \bf II. \ Quantum states in probability space}
\end{center}\vspace{2mm}

A set of  $d^2$ one-dimensional projectors $\{\Pi_i\}_{i=1}^{d^2}$ on a Hilbert space $\mathcal H_d$ is called \emph{symmetric informationally complete}, or \emph{SIC} for short, if
\begin{equation} \label{SIC}
   \textrm{ tr} \,(\Pi_k\,\Pi_l) = \frac{1}{d+1} \ ,\  \ k\neq l \ .
\end{equation}
The value $\frac{1}{d+1}$ is implied by requiring the trace inner product of any pair of projectors to be equal.

\begin{figure} 
\begin{center}
    \includegraphics[scale=0.44]{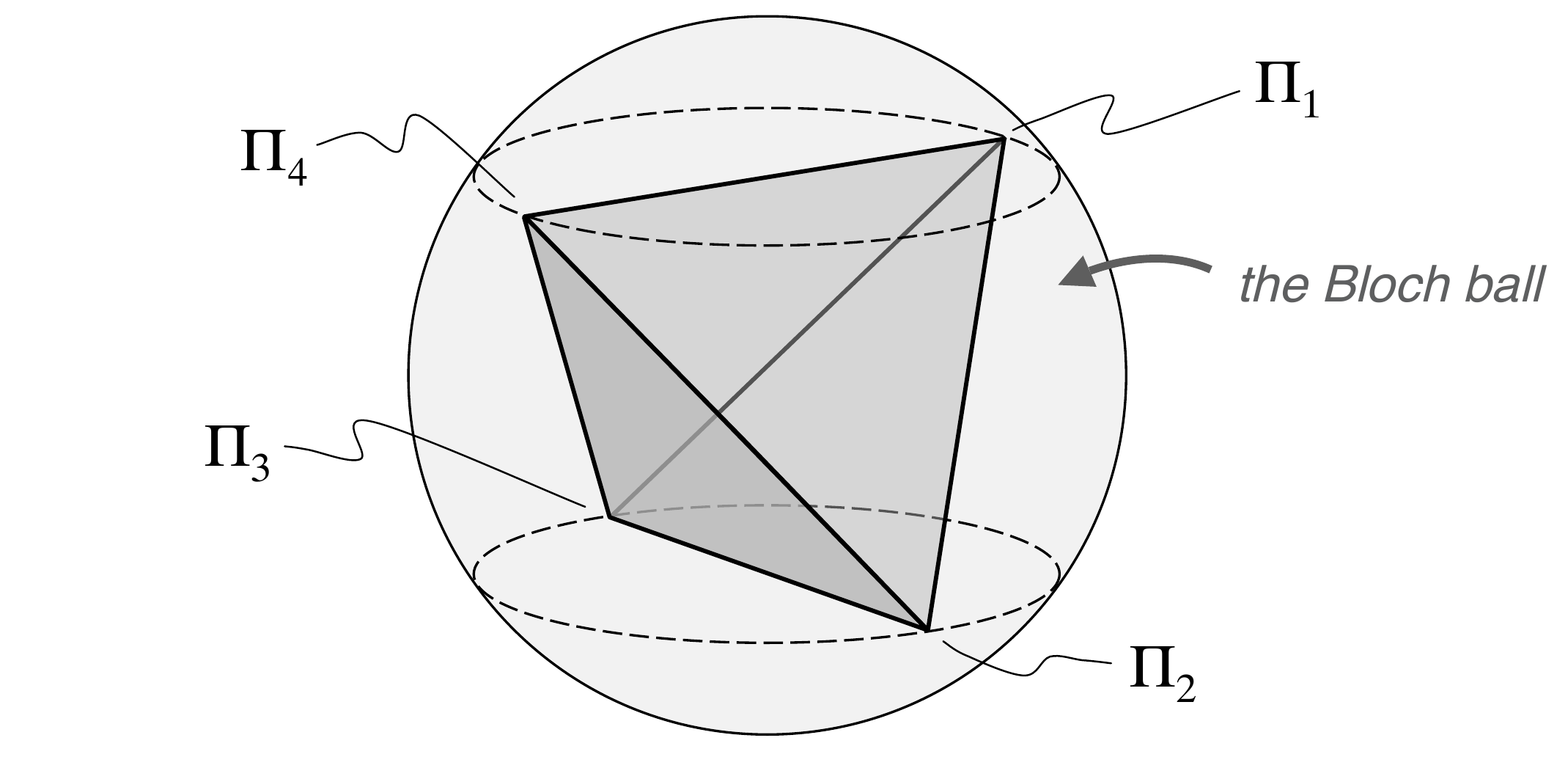}
\vspace{-2mm}
\caption{\small The SIC-projectors when $d=2$ are the corners of a regular tetrahedron inscribed in the Bloch sphere.} \label{tetrahedron}
\vspace*{-4mm}
\end{center}
\end{figure}

The corresponding operators $E_i=\frac{1}{d}\Pi_i$ then form a symmetric informationally complete positive operator-valued measure (POVM)---since $E_i\geq 0$ and it can be shown that $\sum_i E_i=\id\,$---and thus may be seen as an physically possible measurement.  We will use the term SIC for this measurement as well~\cite{Caves99,Renes04}.  It is informationally complete because the operators $E_i$ span the full $d^2$-dimensional space of hermitean operators on $\mathcal H_d$, also a consequence of equation~(\ref{SIC}). And it is called symmetric because the $E_i$ sit at the vertices of a regular simplex in the space of all operators; this is a geometrical reformulation of the equality of the trace inner products in equation~(\ref{SIC}). Moreover the SIC-measurement is minimal since the operators $E_i$ are linearly independent, which entails that no measurement with fewer POVM-elements can be complete. In the simplest case, $d=2$, the four SIC-projectors sit at the vertices of a tetrahedron within the Bloch ball (Fig.~\ref{tetrahedron}).

From the informational completeness of a SIC, it follows that a quantum state, usually thought of as a density matrix~$\rho$, is fully determined by the $d^2$ probabilities for the outcomes of such a measurement,
\begin{equation} \label{probabilities}
  p_i=\textrm{ tr} \,(\rho\,E_i)\;.
\end{equation}
Thus if we choose a fiducial SIC-measurement, every quantum state can be represented as a probability distribution $\mathbf{p}$ in the simplex $\Delta_{d^2}$ of all probability vectors ($p_i\geq 0$ and $\sum p_i=1$) with $d^2$ components. The density matrix is given by
\begin{equation} \label{densitymatrix}
   \rho = \sum_{i=1}^{d^2} \left((d+1)\,p_i - \textstyle{\frac{1}{d}} \right)\Pi_i
           = (d+1)\sum_{i=1}^{d^2} \,p_i \Pi_i - \id \ .
\end{equation}
The coefficients for the SIC-projectors in this expansion are simple functions of the probabilities due to the symmetry of the SIC.

Although every quantum state is uniquely represented by a probability vector, not all probability vectors correspond to quantum states. A simple example is to take $p_1=1$ and the rest $p_i=0$ in equation~(\ref{densitymatrix})---this will not give a positive-semidefinite operator. Thus the density matrices map to a subset of the probability simplex. When the mapping is restricted to be solely between the set of density matrices and this subset, it becomes both one-to-one and onto.

For the last ten years, there has been a lot of research on SICs, and some also earlier under the name of equiangular lines (a selection of references: \cite{Appleby05,Grassl05,Colin05,Scott06,ApplebyDangFuchs}). To define SICs is very simple but to find them is really hard. To this date they have been found numerically only in dimensions $d\leq67$ (though with precision $10^{-38}$). A list of SICs can be found at a web-page by A.~J. Scott~\cite{SICwebpage}.  Furthermore, analytical solutions are known in dimensions $d=2-15$, 19, and 24 \cite{GrasslScott}.  Since the SICs have been found in all dimensions where a serious numerical search has been done, we have significant faith they probably exist for all finite dimensions, and in the following we shall assume so.

Let us turn to the description of quantum-state space as a subset within the probability simplex. Density matrices for pure quantum states are rank-one projection operators; these are the hermitean matrices with $\rho^2=\rho$. Using the expression~(\ref{densitymatrix}) this translates to
\begin{equation}
 p_k=\frac{1}{3}(d+1)\sum_{ij}\alpha_{ijk}\; p_i\,p_ j +\frac{2}{3d(d+1)} \ , \ \ k=1,\dots,d^2 \ ,
\end{equation}
where $\alpha_{ijk}$ are structure constants defined by $\Pi_i \Pi_j=\sum_k\alpha_{ijk}\Pi_k$ (more about the structure constants is in \cite{Fuchs&Schack}; we will not use them here). These $d^2$ coupled quadratic equations determine which probability vectors $\mathbf{p}$ correspond to pure quantum states. An equivalent requirement for pure density matrices is that the operators squared and cubed have unit trace~\cite{Jones05}:
\begin{equation} \label{purecondition}
 \textrm{tr}\, \rho^2 = 1 \ , \qquad
 \textrm{tr}\, \rho^3 = 1 \ .
\end{equation}
The corresponding equations for the probabilities for a SIC-measurement are then:
\begin{equation} \label{extremecondition}
  (i) \quad \sum_i p_i^{\,2} = \frac{2}{d(d+1)} \qquad\quad
  (ii) \quad  \sum_{ijk} \alpha_{ijk} \;p_i\,p_j\,p_k =  \frac{4}{d(d+1)^2}
\end{equation}
This way we get one quadratic equation and one cubic. ($i$) says extreme probabilities, that is, the probability vectors corresponding to extreme elements of the convex set of quantum states, lie on a sphere. ($ii$) includes the structure of the SICs and encodes which part of the sphere actually corresponds to quantum states. The full quantum-state space is the convex hull of the probabilities fulfilling ($i$) and ($ii$).

Since the extreme points lie on a sphere the length of every vector will be less than or equal to the radius, hence the scalar product of any two probability vectors will be bounded:
\begin{equation} \label{upperbound}
 \mathbf{p\cdot q} \leq \frac{2}{d(d+1)}  \  .
\end{equation}
There is equality if and only if $\mathbf{p}=\mathbf{q}$ and $\mathbf p$ is an extreme point. Looking at the trace inner product of a pair of density operators $\rho$ and $\sigma$ in terms of the corresponding probabilities $\mathbf{p}$ and $\mathbf{q}$ we can also get a lower bound on $\mathbf{p\cdot q}$.  That is, since
\begin{equation} \label{tracescalarproduct}
  \textrm{tr}\,\rho\,\sigma = d(d+1)\,\mathbf{p\cdot q} -1 \ ,
\end{equation}
and $\textrm{tr}\,\rho\,\sigma \geq 0$, it follows that
\begin{equation} \label{lowerbound}
  \mathbf{p\cdot q} \geq \frac{1}{d(d+1)} \ .
\end{equation}

In F\&S~\cite{Fuchs&Schack} (and~\cite{Fuchs&SchackII}) it is argued that the Born rule, in its deepest understanding, concerns how to assign probabilities for the outcomes of any possible measurement in terms of the probabilities for the outcomes of a {\it counterfactual\/} two-step measurement---the first step being a SIC measurement that will {\it not\/} actually be performed in the real-world case. Translating the Born rule to SIC-language, one finds that the probability of getting outcome $j$ if one performs a measurement described by POVM-elements $F_j$ is
\begin{equation} \label{bornrule}
  \textrm{Pr}(j) = \sum_i \left(\!(d+1)p_i  - \frac{1}{d} \right) r(j|i) \ ,
\end{equation}
where the conditional probabilities $r(j|i)$ are given by
\begin{equation}
  r(j|i) = \textrm{Tr} \,\Pi_i F_j \ .
\end{equation}
The probability $r(j|i)$ is the probability for obtaining the outcome related to $F_j$ if one first had performed the SIC-measurement and obtained outcome $i$. In the SIC-representation the stochastic matrix with elements $r(j|i)$ uniquely specifies the measurement operators $F_j$. The law of total probability states that the probability of outcome $j$  is $\sum_i p_i r(j|i)$ in the case that the SIC-measurement will first be performed. The Born rule as in equation~(\ref{bornrule}) is the modification needed to take into account that quantum coherence is kept when the SIC-measurement will not be performed. Since probability theory itself says nothing about what the probability $\textrm{Pr}(j)$ (for a performed experiment) will be in terms of $p_i$ and $r(j|i)$ (for an unperformed one), the Born rule can be thought of as an empirical addition to probability theory.

Next in their development, F\&S assume the Born rule in these terms to be a starting point of quantum mechanics---it is taken as a postulate. Probabilities are thus assumed to be calculated according to formula~(\ref{bornrule}), where $\mathbf p$ now stands for the prior for the outcomes of some standard measurement apparatus. The requirement that everything in equation~(\ref{bornrule}) that should be a probability (i.e. $\textrm{Pr}(j)$, $p_i$, and $r(j|i)\,$) actually is a probability (a set of numbers between zero and one and that sum to one) then gives restrictions on what probabilities can correspond to valid quantum states. When this is combined with some assumptions regarding measurements, the inequalities (\ref{upperbound}) and (\ref{lowerbound}) can be re-derived.

The assumptions in short are the following. The probabilities $\mathbf p$ are assumed to span the probability simplex (hence cannot be represented as points in a lower dimensional space). The Principle of Reciprocity has the content that the set of valid priors for the standard measurement is the same as the set of posteriors which one might assign for the standard measurement after having performed any other measurement. Already this requires that the so-called basis distributions---see eq.~(\ref{basisdistribution})---be valid. These are the probabilities we would assign for the outcomes of a SIC-measurement performed on one of the SIC-projectors. The next assumption is more complicated. A measurement is said to have in-step unpredictability (ISU) if the probability assignment for its outcomes is uniform whenever the prior $\mathbf p$ is uniform. Consider the posteriors (from which the uniform distribution is updated) after an ISU measurement with $d$ outcomes has been performed. The assumption is that one of the basis distributions can be obtained as one of the ISU-measurement posteriors, and furthermore that if this had been our prior (as it can be by the Principle of Reciprocity) we would be certain of the outcome of the ISU measurement. For details, further explanations, and motivations, see  F\&S~\cite{Fuchs&Schack}\cite{Fuchs&SchackII}. This leads to the inequalities (\ref{upperbound}) and (\ref{lowerbound}) that we know are true for valid quantum states.

Here we will investigate some implications of these inequalities. The question is, how much of the structure of the set of quantum states is already contained in this constraint?

\begin{figure}[h] 
\begin{center}
    \includegraphics[scale=0.40]{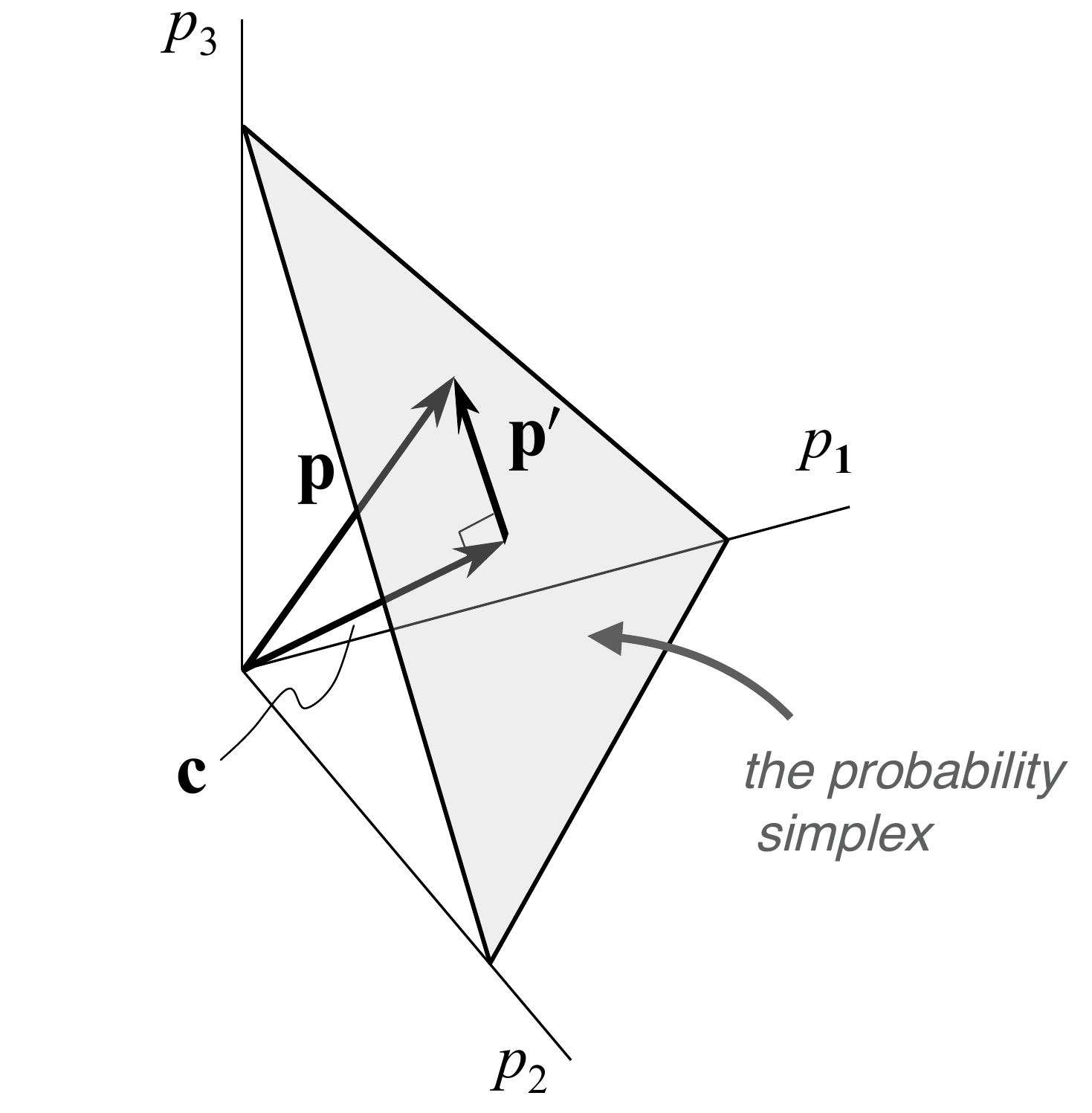}
\vspace*{-2mm}
\caption{\small Sometimes it is convenient to have the uniform distribution $\mathbf c$ as origin. The probability $\mathbf p$ will then be
 represented by the vector $\mathbf p'$.} \label{simplex}
\vspace*{-4mm}
\end{center}
\end{figure}

Before we go on to our study, we note a minor re-expression of probability distributions and the inequalities. It is often helpful to represent points in the simplex by vectors from the midpoint of the simplex rather than as probability vectors $\mathbf{p}$ from the origin in the space coordinatized by outcome probabilities $p_i$. These vectors are parallel to the hyperplane (defined by $\sum p_i=1$) containing the simplex (Fig.~\ref{simplex}). They are given by
\begin{equation}
  \mathbf{p' = p - c} \ , \quad \textrm{where} \quad
  \textstyle{
  \mathbf{c} =\big(\frac{1}{d^2},\dots, \frac{1}{d^2}\big)} \ .
\end{equation}
$\mathbf{c}$ is the midpoint vector, the uniform distribution. Whenever $'$ is used it refers to a vector parallel to the hyperplane. The condition $p_i\geq 0$ for probabilities translates to $p'_i\geq -\frac{1}{d^2}$. With this we get a second version of inequalities~(\ref{upperbound}) and (\ref{lowerbound}),
\begin{equation} \label{primedinequality}
   \frac{-1}{d^2(d+1)} \leq \mathbf{p'\cdot q'} \leq \frac{d-1}{d^2(d+1)} \ .
\end{equation}
%

\begin{center}\vspace{2mm}
  {\Large \bf III. \ QBist convex sets}
\end{center}\vspace{2mm}

We will consider subsets $\mathcal{S}\subset\Delta_{d^2}$  of the probability simplex for which the following holds:
     \begin{itemize}
     \item[(a) ]
$\mathcal{S}$ is \emph{consistent}, which means that for any $\mathbf p, \mathbf q \in \mathcal{S}$
\begin{equation} \label{inequality}
   \frac{1}{d(d+1)} \leq \mathbf{p\cdot q} \leq \frac{2}{d(d+1)} \ .
\end{equation}

     \item[(b) ]
$\mathcal{S}$ is \emph{maximal} in the sense that adding any further $\mathbf p\in\Delta_{d^2}$ to $\mathcal{S}$ makes it inconsistent.

     \item[(c) ]
$\mathcal{S}$ contains $d^2$ probability vectors called \emph{basis distributions}; these are
\begin{equation} \label{basisdistribution}
\textstyle{
  \mathbf{e_k}=\big(\frac{1}{d(d+1)},\dots,\frac{1}{d(d+1)},\frac{1}{d},\frac{1}{d(d+1)},\dots,\frac{1}{d(d+1)}\big)}
    \ , \ \ \mathbf{k}=1,\dots,d^2 \, ,
\end{equation}
with the larger value $\frac{1}{d}$ in the $k$th postition.

     \item[(d) ]
Every $\mathbf p$ for which the upper bound in (a) is attained when $\mathbf{q=p}$ belongs to at least one set $\{\mathbf{p_k}\}$ with a maximum number $m$ of \emph{maximally distant points} allowed by (a). This means that the lower bound in (a) is attained for every distinct pair $\mathbf{p_k}$ and $\mathbf{p_l}$.  All together,
\begin{equation} \label{maxdistant}
  \mathbf{p_k\cdot p_l}= \frac{\delta_\mathbf{kl}+1}{d(d+1)} \ , \ \ \mathbf{k,l}=1,\dots,m \ .
\end{equation}
     \end{itemize}

Sets $\mathcal{S}$ satisfying these criteria in will be called {\it QBist state spaces}.

\vspace{2mm}

\noindent
\emph{Remarks:}\\[3pt]  
---~(a)~is, as stated in the previous section, a consequence of taking the Born rule as it appears in SIC-language as a postulate together with a few other assumptions. Instead of the bounds~(\ref{inequality}) in (a) we frequently use the equivalent bounds given in equation~(\ref{primedinequality}) for vectors in the hyperplane of the simplex.\\[3pt]
---~(b)~is reasonable if we want to accept all probability assignments not ruled out by the other conditions.  Note however, that as stated it is not clear that a set which is maximal can fulfill also (c) and~(d). (For example, it might be that the extension to a maximal set will always violate~(d), although we have seen no indication to something like that).\\[3pt]
---~(c)~might seem an ad hoc assumption at first sight, but the validity of these probabilities, that is that they should be included in the state space, is, as mentioned in the previous section, a natural fallout of the analysis by F\&S~\cite{Fuchs&Schack} leading to inequalities~(a).\\[3pt]
---~(d)~introduces a maximum number $m$, which will be shown to be equal~$d$. This criterion is a way to ``spread out'' the set $\mathcal S$ as much as possible and at the same time ascertain an equal footing for all probabilities that are extreme in the sense that $\mathbf{p\cdot p}$ attains the upper bound in~(a). This requirement seems rather strong and we guess it might be crucial for ultimately re-obtaining quantum-state space.  Nonetheless, it will play a minor role in the investigations of the present paper.

\vspace{2mm}

We will see that already the seemingly simple requirements (a) and (b) of consistency and maximality lead to nontrivial features for the sets~$\mathcal S$, some of which are discernable also for the set of quantum states. When we further impose (c) and~(d), about basis distributions and maximally distant points, we know yet only of one example of such a set~$\mathcal S$, the actual quantum-state space.

\subsubsection{Maximally consistent means quantum when $d=2$} 

\indent \indent
We first have a quite detailed look at the special case when $d=2$. From (a) in the form of equation~(\ref{primedinequality}) we get
\begin{equation}
  \frac{-1}{12} \leq \mathbf{p'\cdot q'} \leq \frac{1}{12} \ .
\end{equation}
The upper bound gives a sphere (with radius $1/\sqrt{12}$). The lower bound doesn't give any constraint since even two ``opposite'' vectors on the sphere (as if $\mathbf{q'=-p'}$) do not give a smaller scalar product.  Thus, because of (b) every $\mathbf p'$ within this sphere should be included in $\mathcal S$. Now we need to ascertain that all vectors in the sphere are probabilities, that is, that the whole sphere lies within the probability simplex, in this case a tetrahedron. Well, the probability $\mathbf p =(\frac{1}{3},\frac{1}{3},\frac{1}{3},0)$ is one of the boundary points of the tetrahedron being closest to the midpoint $\mathbf c$---it is the midpoint of a facet---and the distance is $1/\sqrt{12}$. Since this is the same as the radius of our sphere, we see that it actually is the insphere of the tetrahedron (Fig.~\ref{insphere}).

(c) requires vectors like $\mathbf{e_1}=(\frac{1}{2},\frac{1}{6},\frac{1}{6},\frac{1}{6})$ to belong to $S$. They are on the sphere since  $\mathbf{e_1\cdot e_1}=\frac{1}{3}$ is just the upper bound in (a). The maximal distance is for antipodal points on the sphere, $\mathbf p$ and $-\mathbf p$, so the maximum number of maximally distant points is $m=2$. Obviously every point on the sphere is in a set of two antipodal points and thereby (d) is satisfied too.

\begin{figure}[h] 
\begin{center}
    \includegraphics[scale=0.44]{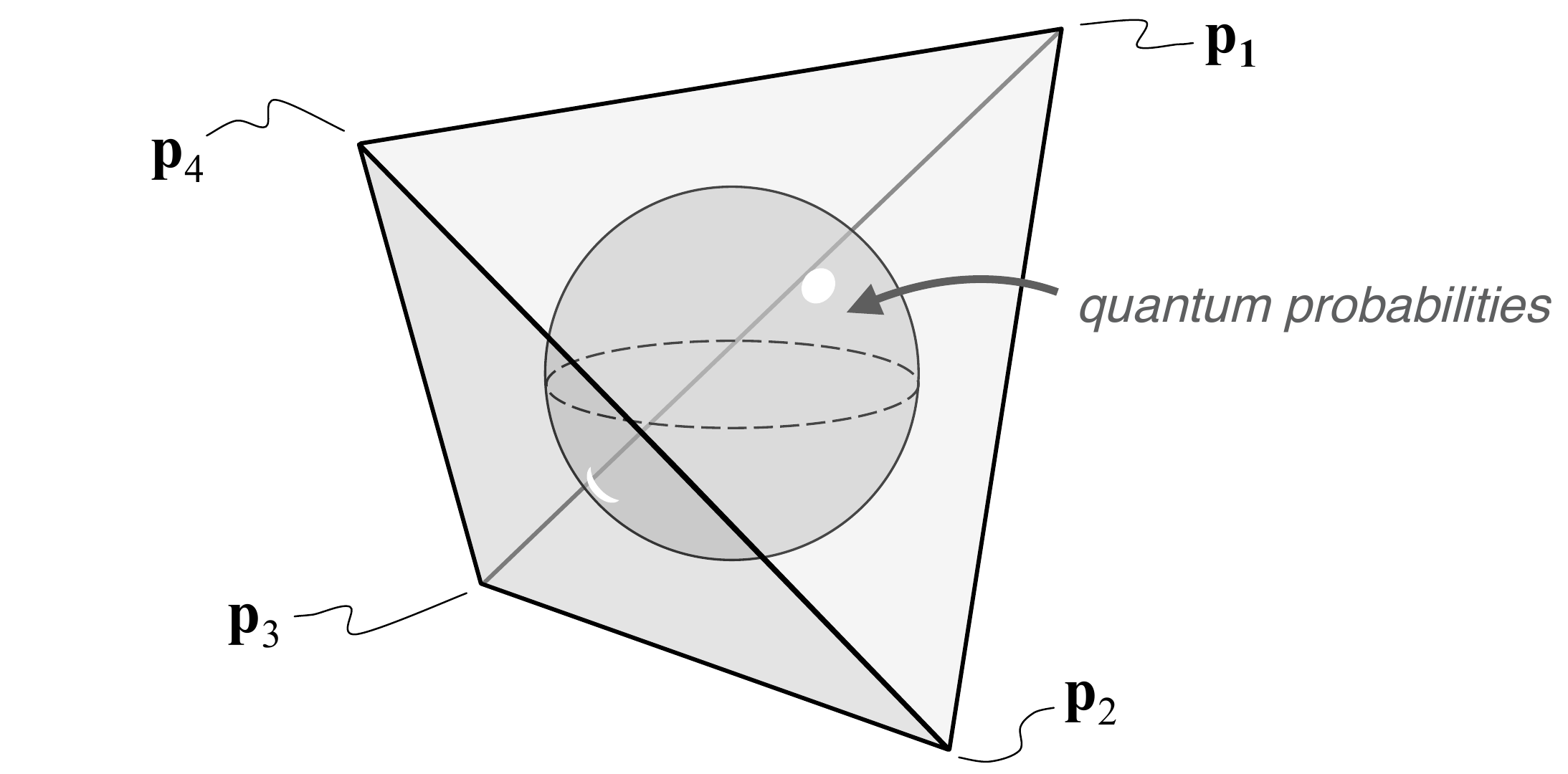}
\vspace*{-2mm}
\caption{\small When $d=2$ quantum-state space is the insphere of the probability simplex. It is fully characterized by maximality and consistency.} 
\vspace*{-4mm}
\end{center}
\end{figure}

This unique maximally consistent set $\mathcal S$ is actual quantum-state space in SIC representation. When $d=2$, pure density matrices are determined by $\textrm{tr}\,\rho^2=1$, which is equivalent to the sphere condition. The more complicated equation $(ii)$ in (\ref{extremecondition}) is automatically satisfied in this case; it only becomes nonredundant in $d\ge 3$.

\subsubsection{Convexity, compactness and strict bounds} 

\indent \indent
All QBist state spaces $\mathcal{S}$ are \emph{convex}. For, if $\mathbf p_1$ and $\mathbf p_2$ are in the set, so is $\mathbf p=\lambda \mathbf p_1+(1-\lambda) \mathbf p_1$, $0<\lambda<1$. This follows from the linearity of the scalar product and the maximality of~$\mathcal{S}$.

$\mathcal{S}$ is bounded by the sphere centered at the midpoint of the simplex with radius~$r'$, determined by equation~(\ref{primedinequality}) to be given by
\begin{equation} \label{radius}
    r'^{\,2} = \frac{d-1}{d^2(d+1)} \ ,
\end{equation}
Because of condition (c) it follows that this is the minimal circumscribed sphere of $\mathcal{S}$. The basis distributions $\mathbf{e_k}$  lie on the sphere and they form a regular $(d^2-1)$-dimensional simplex (since they are all at the same distance from each other) inscribed in the sphere. Later we show that this sphere is not completely contained within the probability simplex for $d>2$.

Because the inequalities in (a) are not strict, $\mathcal{S}$ is not only bounded but also closed and thus compact. For, consider any point $\bar{\mathbf p}\in\bar{\mathcal S}\,$, $\bar{\mathcal S}$ being the closure of $\mathcal S$. By definition there is a sequence $\mathbf{p_t}$ in $\mathcal S$ converging to  $\bar{\mathbf p}$. For each $\mathbf{p_t}$ in the sequence and all $\mathbf q \in \mathcal S$ it holds that
\begin{equation}
   \frac{1}{d(d+1)} \leq \mathbf{p_t\cdot q} \leq \frac{2}{d(d+1)} \ .
\end{equation}
This inequality will then hold also for the limit point $\bar{\mathbf p}$ and we have by maximality that $\mathcal S=\bar{\mathcal S}$.

Using the terminology of convex sets, $\mathcal{S}$ consists of two distinct subsets: the extreme points (vectors that cannot be obtained as a convex combination of any others in $\mathcal{S}$) and the mixed points (convex combinations of extreme ones). Especially every point in $\mathcal{S}$ on the sphere will be extreme. In the language of quantum mechanics extreme points are pure states.

From these geometrical considerations one can see that the upper bound in (a) is strict,
\begin{equation}
   \frac{1}{d(d+1)} \leq \mathbf{p\cdot q} < \frac{2}{d(d+1)} \ ,
\end{equation}
whenever $\mathbf q \neq \mathbf p$, or when $\mathbf{p=q}$ but $\mathbf{p}$ is interior to the sphere. We can also prove it algebraically by contradiction: Assume there is equality in the upper bound for~$\mathbf{p\cdot q}$. This would give
\begin{equation} \label{proof}
   \sum_i (p_i-q_i)^2 =  \mathbf{p\cdot p} - 2\,\mathbf{p\cdot q} +\mathbf{q\cdot q} \leq
       \frac{2}{d(d+1)}-2\,\frac{2}{d(d+1)}+\frac{2}{d(d+1)}=0 \ ,
\end{equation}
where the inequality comes from the terms $ \mathbf{p\cdot p}$ and $\mathbf{q\cdot q}$. If there is a strict inequality in the middle this line is just false, thus we have a contradiction to the assumption that $\mathbf{p\cdot q}$ attains the upper bound. In the case of equality in the middle---that is, if both $\mathbf p$ and $\mathbf q$ are on the sphere---the conclusion is that $\mathbf{p=q}$.

\subsubsection{Maximally distant points} 

\indent \indent
Consider probability distributions $\mathbf{p_k}$ and $\mathbf{p_l}$ in $\mathcal{S}$ that fulfill equation~(\ref{maxdistant}) in condition~(d). They are points on the sphere such that the vectors $\mathbf{p_k}$ and $\mathbf{p_l}$ are ``as orthogonal as possible,'' without violating (a). Or, when considering the vectors $\mathbf{p'_k}$ and $\mathbf{p'_l}$, they are as anti-parallel as possible. Criterion (d) includes a maximum number $m$ of such maximally distant points. In F\&S~\cite{Fuchs&Schack} this value was shown to be at most $d$ by means of considering a Gram matrix. Here we give a more elementary argument for this bound.

The set of $m$ maximally distant points will form a regular simplex $\Delta_m$. The opening angle $\delta_m$ between the lines from the midpoint of such an $(m-1)$-dimensional simplex to the vertices is $\cos\delta_m=\frac{-1}{m-1}$. Now assume the bounds in equation~(\ref{primedinequality}) are attained:
\begin{equation}
  \mathbf{p'_k\cdot p'_k} = \mathbf{p'_l\cdot p'_l} = \frac{d-1}{d^2(d+1)}
  \ , \qquad
  \mathbf{p'_k\cdot p'_l} = \frac{-1}{d^2(d+1)} \ .
\end{equation}
Let $\theta$ be the angle between $\mathbf{p'_k}$ and $\mathbf{p'_l}$ and combine the equations above:
\begin{equation}
  \mathbf{p'_k\cdot p'_l} = \mathbf{|p'_k| |p'_l|} \cos\theta = \mathbf{|p'_k|^2} \cos\theta =
  \mathbf{p'_k\cdot p'_k}  \cos\theta
\end{equation}
\begin{equation}
  \Rightarrow \qquad
   \cos\theta = \frac{\mathbf{p'_k\cdot p'_l}}{\mathbf{p'_k\cdot p'_k}} = \frac{-1}{d-1}
\end{equation}
This is the same as for the opening angle of a $(d-1)$-dimensional simplex $\Delta_d$. Thus we can choose maximally distant points to form a simplex centered at the midpoint of the probability simplex $\Delta_{d^2}$, or equivalently the midpoint of the sphere, and then there will be $d$ maximally distant points. One can think of the vertices of this lower dimensional simplex $\Delta_d$ as situated at a $(d-1)$-dimensional ``equator'' of the $(d^2-1)$-dimensional sphere. Any set of maximally distant points not centered at the midpoint of the sphere will have fewer than $d$ vertices.

That there can be no more than $d$ maximally distant points is clearly seen if we calculate the length of the vector $\mathbf G'=\sum_{k=1}^m \mathbf{p'_k\,}$, which of course has to be positive:
\begin{equation}
  \mathbf{G'\cdot G'} = \sum_{k,l=1}^m  \mathbf{p'_k\cdot p'_l} =
  \sum_{k,l=1}^m \frac{d\delta_\mathbf{kl}-1}{d^2(d+1)} = \frac{m(d-m)}{d^2(d+1)} \ .
\end{equation}

These considerations also show that the uniform mixture---the convex combinations with equal weights---of a set of $d$ maximally distant probabilities has to be the uniform probability distribution $\mathbf c\,$:
\begin{equation}
  |\mathbf G'| = 0     \qquad \Leftrightarrow \qquad
  \sum_{k=1}^d \frac{1}{d}\,\mathbf{p'_k} = 0     \qquad \Leftrightarrow \qquad
  \sum_{k=1}^d \frac{1}{d}\,\mathbf{p_k} = \mathbf c
\end{equation}

Note that the above argument makes no reference to the nonnegativity of the components of the $\mathbf{p_k}$.  That is, it was shown that there can be at most $d$ maximally distant points on the sphere, but can these points all really be probabilities?  Or, would making explicit use of the constraint that all the components of all the vectors be nonnegative force a tighter upper bound that is something below $d$.  As will be seen in a later section, some parts of our sphere are actually outside the probability simplex. That is, some parts of the sphere are not elements of $\mathcal S$. So this is not a trivial question.  However, that it is possible to orientate a simplex $\Delta_d$ so that it is contained in the probability simplex $\Delta_{d^2}$ can be seen by invoking quantum mechanics. An orthogonal basis in Hilbert space $\mathcal H_d$ corresponds, via the SIC-representation, to a set of $d$ maximally distant points (see equations (\ref{tracescalarproduct}) and~(\ref{lowerbound})). It would be nice not to have to refer to quantum mechanics and SICs to show that our bound is tight, but so far a direct proof eludes us.

\subsubsection{Maximal-valued probabilities} 

\indent \indent
A vector like $\mathbf p=(1,0,\dots,0)$ is outside the sphere from condition (a), so we cannot have an outcome probability $p_i=1$ for any vector in $\mathcal S$. What then, is the largest probability $p_i$ allowed? That the maximal value is $p_i=\frac{1}{d}$ can be seen in several ways. In F\&S~\cite{Fuchs&Schack} it is shown by looking at the scalar product with the basis distributions, included in $\mathcal S$ according to condition (c). We have
\begin{equation}
   \mathbf{p\cdot e_k} = \frac{1}{d(d+1)}+ p_k \frac{1}{(d+1)}  \ , \quad \mathbf{k}=1,\dots,d^2 \, .
\end{equation}
This is greater than the bound in (a) unless $p_k\leq \frac{1}{d}$ for all $k$. So the basis distributions are examples---in fact the only ones---of probability vectors with a maximal-valued element.

Even if we do not assume (c), that is that the basis distributions have to be included in $\mathcal S$, the same bound holds. This follows because $\mathbf{e_1}$ is a probability vector on the sphere that bounds $\mathcal S$, whether it is in $\mathcal S$ or not. Moreover $\mathbf{e_1}$ has one large component $p_1= \frac{1}{d}$ with the rest being equal and smaller. From symmetry it must be the point on the sphere closest to the vertex $(1,0,\dots,0)$ of the probability simplex. Hence no other point on the sphere or within it---in particular no point in~$\mathcal S$---can have a larger probability than $ \frac{1}{d}$ for the first outcome, or of course for any other outcome either.

A formalized version of the argument goes like this. Start with the sphere condition, assume $p_1$ is maximal-valued, use a variant of the Schwarz inequality, and that the probabilities sum to one:
\begin{equation}
  {\textstyle \frac{2}{d(d+1)}} = \sum_{i=1}^{d^2} p_i^2 = p_1^2 +  \sum_{i=2}^{d^2} p_i^2 \geq
    p_1^2 + {\textstyle  \frac{1}{d^2-1}}\Big(\sum_{i=2}^{d^2} p_i\Big)^2 =
    p_1^2 + {\textstyle \frac{1}{d^2-1}}\big(1-p_1\big)^2
\end{equation}
A little algebra and this gives $p_1\leq\frac{1}{d}$, where there is equality if all the other $p_i$ are equal.

The quantum density operators corresponding to the basis distributions $\mathbf{e_k}$ are the SIC-projectors $\Pi_k$ themselves. That probabilities are bounded is a general feature of informationally complete measurements, since the POVM-elements are then not orthogonal.

\subsubsection{Zero-probabilities and broken symmetry} 

\indent \indent
We have seen that the extreme value $p_i=1$ is not possible, but what about probability vectors with some probabilities $p_i=0$? If any component of a vector $\mathbf p$ is zero this means that it lies on the boundary of the probability simplex. If there are $n$ zero-probabilities $p_i=0$, then $\mathbf p$ is a point in a face of dimension $d^2-n-1$. The question of how many zero-probabilities are possible is thus a question of how far out the sphere reaches, which faces it intersects.

The point in a  $(d^2-n-1)$-dimensional face that is closest to the midpoint $\mathbf c=(\frac{1}{d^2},\cdots\frac{1}{d^2})$ of the simplex is $(\frac{1}{d^2-n},\cdots,\frac{1}{d^2-n},0,\dots,0)$, or some permutation thereof, at a distance $d_\textrm{\scriptsize face}$ given by
\begin{displaymath}
  d_\textrm{\scriptsize face}^2 =
  (d^2-n)\Big(\frac{1}{d^2}-\frac{1}{d^2-n}\Big)^2+n\Big(\frac{1}{d^2-0}\Big)^2=\frac{n}{d^2(d^2-n)} \ .
\end{displaymath}
This should be compared to the radius $r'$ of the sphere given by equation~(\ref{radius}). Requiring $r'^2\geq d_\textrm{\scriptsize  face}^2$ gives the bound
\begin{equation} \label{zeroesbound}
  n\leq \frac{1}{2}d(d-1)
\end{equation}
for the maximal number of zero-probabilities. Furthermore, an equality means that the sphere just touches the face at its midpoint, that is, a point like $\mathbf p=(\frac{2}{d(d+1)},\cdots,\frac{2}{d(d+1)},0,\dots,0)$. Up to permutations, this is the only probability vector on the sphere having $\frac{1}{2}d(d-1)$ zeroes, and hence $\frac{1}{2}d(d+1)$ nonzero probabilities.

\begin{figure}[b] 
\begin{center}
    \includegraphics[scale=0.44]{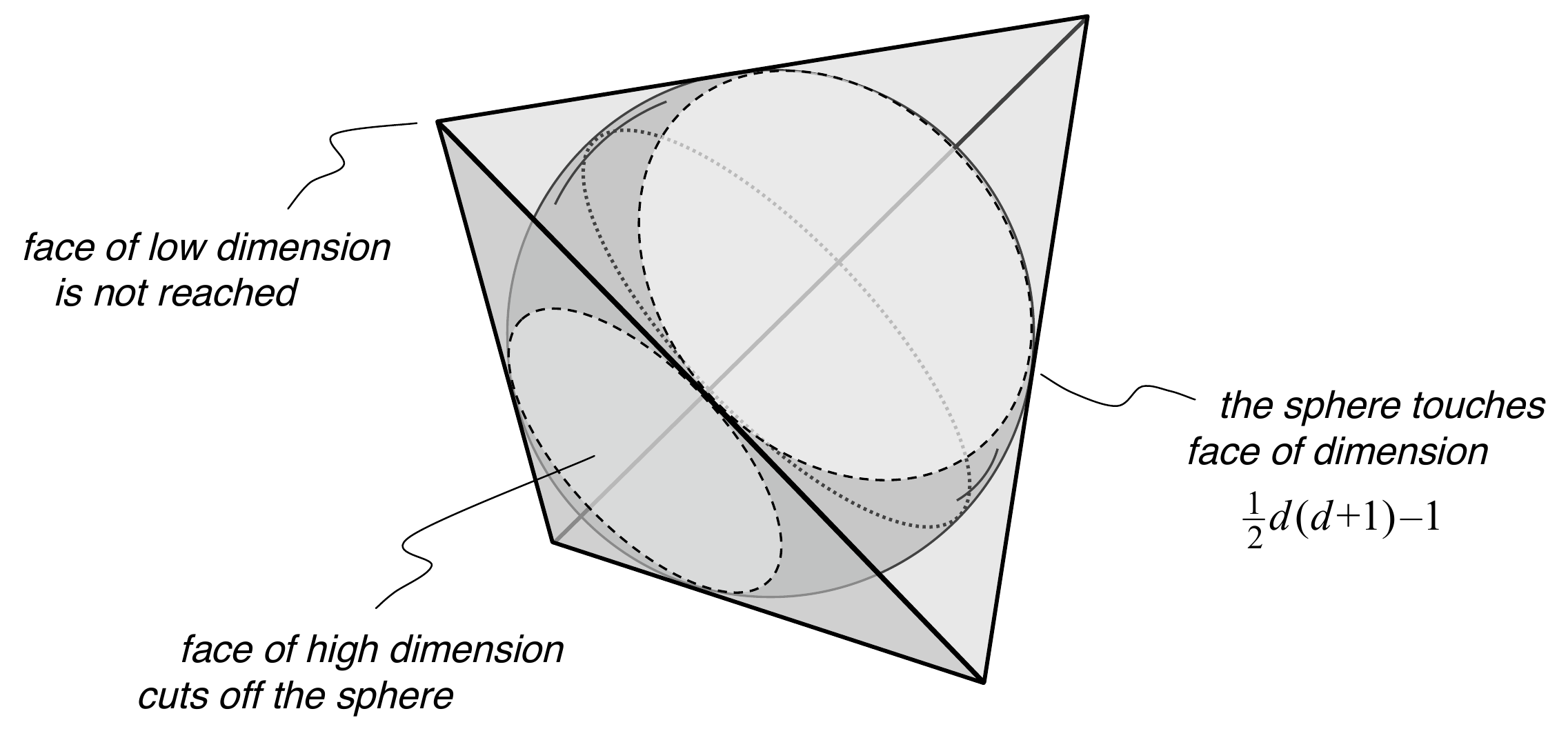}
\vspace*{-2mm}
\caption{\small A schematic picture of how the sphere constraining the set $\mathcal S$ relates to different faces of the probability simplex. (Note that this only indicates the situation in higher dimensions; when $d=2$ the sphere is inscribed in the tetrahedron.)} \label{faces}
\vspace*{-4mm}
\end{center}
\end{figure}

Similar to our alternative for finding the maximal probability via the Schwarz inequality, this bound on the maximal number of zero-probabilities can be obtained from the following line of reasoning:
\begin{equation}
  1 = \Big(\!\sum_{\{i|p_i\neq0\}} \! p_i\Big)^2 \leq
  (d^2-n)\!\!\!\sum_{\{i|p_i\neq0\}}\! p_i^2 \leq (d^2-n)\,\frac{2}{d(d+1)} \ .
\end{equation}

We have seen that the sphere reaches to faces of dimension $\frac{1}{2}d(d+1)-1$. Lower dimensional faces of the probability simplex lie fully on the outside of the sphere, whereas the sphere pokes out of the simplex through the higher dimensional faces (unless $d=2$ which was studied earlier). So some parts of the sphere have to be excluded from the set~$\mathcal S$, since those points do not correspond to probabilities (Fig.~\ref{faces}). Furthermore, because of the lower bound in (a), still more points have to be excluded:  As we will see, it means that at least some vectors with $n=\frac{1}{2}d(d-1)$ zeroes cannot be included in~$\mathcal S$.

Assume the vector
\begin{equation} \label{zeroesprobabilities}
  \mathbf p = \frac{2}{d(d+1)}
       (\,\underbrace{1\,,\,.\dots.\;,1}_{\frac{1}{2}d(d+1)}\, ,
        \,\underbrace{0\,,\,\dots\,,0}_{\frac{1}{2}d(d-1)}\,)
\end{equation}
is in the set $\mathcal S$. This is compatible with criteria (a)--(c)---especially it can be verified that the scalar product, $\mathbf{p\cdot e_k}$, with the basis distributions are within the limits of~(a). For each $\mathbf{e_k}$ with $\mathbf{k}>\frac{1}{2}d(d+1)$, this scalar product attains the minimum, hence $\mathbf p$ is maximally distant to those~$\mathbf{e_k}$. Whether (d) can also be fulfilled is too early to say.

If $\mathbf p$ is allowed, it might seem reasonable that the vectors obtained from permuting the components of $\mathbf p$, for example
\begin{equation}
  \mathbf{\tilde p} = \frac{2}{d(d+1)}
        (\,0\,,\,\dots\,,0\, , \,1\,,\,.\dots.\;,1\,) \ ,
\end{equation}
should also be allowed. But $\mathbf{p\cdot\tilde p}$ is too small, they are too far apart, whenever $d>3$. Those vectors valid together with $\mathbf p$ are permutations where the number of nonzero components not altered is at least
\begin{equation}
  s_1 \geq \frac{1}{4}d(d+1) \ .
\end{equation}
The upper bound $\mathbf{p\cdot p}=\frac{2}{d(d+1)}$ is obtained when all nonzero components ``overlap'' and the lower bound  $\mathbf{p\cdot p_\sigma}=\frac{1}{d(d+1)}$ when half of the nonzero components ``overlap'' after some suitable permutation $\sigma$. This is clearly only possible when $d$ or $d+1$ is divisible by 4; for other dimensions consistent pairs of vectors of this type is not maximally distant. The number of zero components in same position will be $s_0 \geq \frac{1}{4}d(d-3)$, and, to play around with numbers a little more, the total number of components in the same position has to be $s \geq \frac{1}{2}d(d-1)$.

The above is a straightforward consequence of the limits in criteria (a), but it is still rather surprising. We are studying subsets $\mathcal S$ of the probability simplex $\Delta_{d^2}$, and we are requiring that the $d^2$ basis distributions in (d), symmetrically positioned relative to the probability simplex, is in our set. Despite this it turns out the symmetry of the set is not the permutational symmetry of the simplex. This broken symmetry does not hinge crucially on including one of these max-zeroes vectors, the situation is similar for other vectors.

If a quantum state $\mathbf p$ in the SIC-representation contains zero-probabilities $p_i=0\,$, this is equivalent to the corresponding density matrix $\rho$ being orthogonal to the SIC-projectors $\Pi_i$ (from equation~(\ref{probabilities})). Consider the SIC-vectors in Hilbert space $\mathcal H_d$. Take any $d-1$ of them and they will for sure lie in a $(d-1)$-dimensional subspace (or lower if they are linearly dependent). Then there exists a pure state $|\psi\rangle$ orthogonal to these chosen $d-1$ SIC-vectors, and the corresponding SIC-representation  $\mathbf p$ will have zeroes for these components. So from quantum mechanics we see that $d-1$ zero-probabilities $p_i=0$ will for sure be possible; this is a lower bound on the upper bound on the maximal number of zeroes.

Finding an upper bound within quantum mechanics boils down to the question of how many of all $d^2$ SIC-vectors can be confined into a $d-1$ dimensional subspace of $\mathcal H_d$. The best known general bound is $\frac{1}{2}d(d-1)$, which is the same as the one we found for any consistent set $\mathcal S$.

\begin{figure} 
\begin{center}
    \includegraphics[scale=0.44]{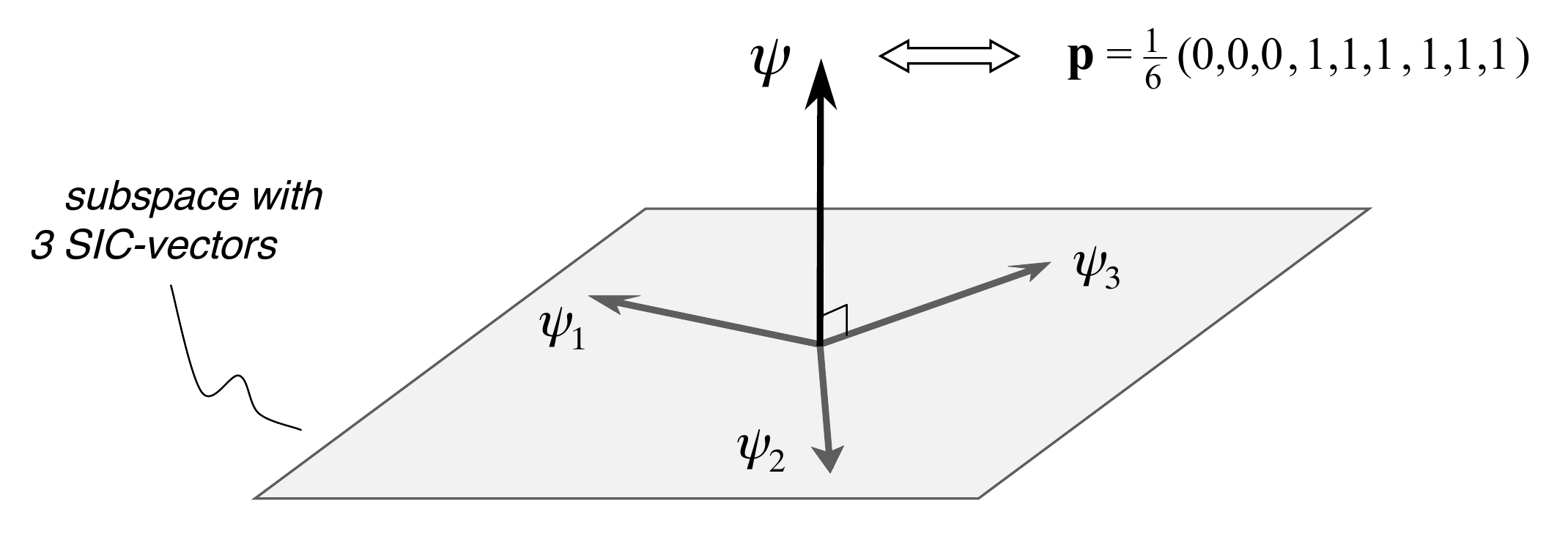}
\vspace*{-4mm}
\caption{\small A state vector which is orthogonal to three linearly dependent SIC-vectors in $\mathcal H_3$ and its corresponding probability representation.} \label{zeroesvector}
\vspace*{-2mm}
\end{center}
\end{figure}

In dimension $d=3$ no permutation of $\frac{1}{6}(1,1,1\,,\,1,1,1\,,\,0,0,0)$ will be excluded from the pairwise scalar product. When no zeroes ``overlap'' the lower bound in (a) will be attained and thus the points are maximally distant. But not all those vectors are quantum states in a SIC-representation. Dimension 3 is special in that there is a one-parameter family of distinct SICs known. Some of these have the property that for any two SIC-vectors chosen there is a third one linearly dependent~(Fig.~\ref{zeroesvector}). Equivalently, for any two $p_i=0$ there is a third one which can also be zero. Also for some of the other SICs one can find three SIC-vectors in a subspace, thus the potentiality of having three $p_i=0$, but there are not as many possible combinations. Still other SICs might not allow more than two zeroes.

In dimensions $d=4$ and $d=5$ exhaustive searches for the known SIC-sets show that there is no set of more than $d-1$, that is $3$ respectively $4$, SIC-vectors in a $d-1$ dimensional subspace. But in dimension 6 there is again sets of $d$ vectors confined in a $d-1$ dimensional subspace. We do not know what the situation is for higher dimensions.

In dimension $d=4$ the exclusion of some permuted probability vectors might be more forceful when considered together with condition (d). Start with a vector of the form~(\ref{zeroesprobabilities}). To find vectors $\mathbf p_\sigma$ maximally distant, and also pairwise maximally distant, is a combinatorial task. By trial and error we found that there can be no more than three, which is less than the $m=4$ bound on maximally distant points. Perhaps this is a clue that no probabilities with as many as 6 zeroes occur for quantum states.

\begin{center}
  {\Large \bf IV. \ Discussion}
\end{center}

\noindent
We have imposed four criteria (a)--(d) on a set $\mathcal S$ that is a subset of the probability simplex $\Delta_{d^2}$. In short they are as follows:\\[4pt]
\hspace*{1pt} (a) places upper and lower bounds on scalar products---the set should be \emph{consistent}.\\
\hspace*{1pt} (b) requires the set to be \emph{maximal} (no more point can be added consistently).\\
\hspace*{1pt} (c) includes $d^2$ \emph{basis distributions}.\\
\hspace*{1pt} (d) is a condition regarding \emph{maximally distant points}.\\[4pt]
To some degree these have been motivated by F\&S~\cite{Fuchs&Schack} when they, in their advancement of Quantum Bayesianism, propose that the Born rule seen as an empirical addition to probability theory is the primal law of quantum mechanics. Our aim here has been to begin investigating what one can say about this type of set~$\mathcal S$, and to see how close examples of the type are to quantum-state space (in SIC-representation).

In dimension 2 the only maximally consistent set is quantum-state space itself. In higher dimensions we can think of other maximally consistent sets, but we do not know of any other than quantum-state space that fulfills all the conditions (a)--(d).

As an example of another maximally consistent set, start with a small cap of the circumscribed sphere, that is, everything in the $(d^2-2)$-dimensional spherical hyper-surface within a solid angle which is not too large so that the bounds in (a) are fulfilled. Extend this set by adding points consistently until it is maximal. Such a set is not quantum-state space. It cannot be ``quantum'' since every point in the cap is extreme and this set of extreme points is thus of too high a dimension. A set built in this way will violate condition (d) about maximally distant points. For, consider a point somewhere in the interior of the cap and another point maximally distant from the first. This second point will be too far away from some points in the neighborhood of the first (unless the $d=2$, when maximally distant points are antipodal points).

The argument above shows that the requirement of maximally distant points constrains the dimension of the set of extreme points (at least those in the circumscribed sphere). All extreme points of quantum-state space---the pure states---lie on the circumscribed sphere and form a connected set of dimension $2d-2$. Is it possible to deduce these properties from the criteria (a)--(d)? We hope future work will give an answer. An additional criterion could be that all extreme states must lie on the circumscribed sphere, that is, attaining the upper bound in (a). Yet it might not be needed. If the set $\mathcal S$ could be confined to one end of the probability simplex there would be extreme points not on the sphere, but this cannot be the case because the basis distributions in (c) are spread out in all directions.

When discussing extreme states and sets of maximally distant states it is worth mentioning a very special property of quantum-state space. Any quantum state, although in $(d^2-1)$ dimensions, can be written as a convex combination of no more than $d$ extreme states. Furthermore, these extreme states form a set of maximally distant points (the probability distributions corresponding to an orthogonal basis in Hilbert space). For a general point in a general convex set of the same dimension one would need a convex combination $d^2$ extreme points by Carath\'eodory's theorem.  Thus quantum-state space holds a unique position with respect to some very basic convexity properties.  We wish we knew whether one could prove such a tighter bound for QBist state spaces.

Quantum state space in the SIC-representation is determined by two equations~(\ref{extremecondition}): (i) asserts that extreme states lie on a sphere, whereas (ii) is more complicated and includes structure constants $\alpha_{ijk}$. These hold the details of the actual SIC, and indirectly the structure of the unitary group SU($d$). This is what we want to characterize without referring to a given SIC. Still it can of course only be achieved under the assumption that SICs exist.

The hope is that quantum-state space can be characterized by something similar to the criteria (a)--(d) above, perhaps only with minor extensions. Although still far from a proof of such a characterization, we have demonstrated that all QBist state spaces display several nontrivial similarities with quantum-state space. Furthermore, we know quantum-state space is a QBist state space, and so far it is the only example we know.

\begin{center}
 \textbf{Acknowledgements}
\end{center}

We thank Matthew Graydon for sharing his example of 6 linearly dependent SIC-vectors in dimension 6. This research was supported in part by the U.~S. Office of Naval Research (Grant No.~N00014-09-1-0247). Research at Perimeter Institute is supported by the Government of Canada through Industry Canada and by the Province of Ontario through the Ministry of Research \& Innovation. \AA.~E.~acknowledge support from The Wenner-Gren Foundations.

\end{document}